\documentstyle[prc,aps,twocolumn,psfig]{revtex}

\newcommand{\be}{\begin{equation}}
\newcommand{\ee}{\end{equation}}
\newcommand{\bea}{\begin{eqnarray}}
\newcommand{\eea}{\end{eqnarray}}

\newcommand{\FF}{\Phi_0^2+\sum_i\Phi_i^2}

\begin{document}
\title{Dissipation and Fluctuation
at the Chiral Phase Transition}

\author{{\sc Tam\'as S. Bir\'o$^{\, a}$ 
and Carsten Greiner$^{\, b}$},\\
$^{a}$
{ \it MTA KFKI RMKI Theory Division,
H-1525 Budapest, Hungary, } \\
$^{b}$ { \it Institut f\"ur Theoretische Physik, Universit\"at Giessen,
D-35392 Giessen, Germany. } }

\date{April 2, 1997}

\maketitle
%\thispagestyle{empty}
%\renewcommand{\thefootnote}{\arabic{footnote}}

%%%%%%%%%%%%%%%%%%%%% ABSTRACT %%%%%%%%%%%%%%%%%%%%%%%%%%%%

\begin{abstract}
Utilizing the Langevin equation for the linear $\sigma$-model we
investigate the effect of friction and  white noise on the 
evolution and stability
of collective pionic fields in energetic heavy ion collisions.
We find that the smaller the volume, the more stable transverse
(pionic) fluctuations become on a homogeneous disoriented chiral
field background (the average transverse mass $\langle m_t^2\rangle$ increases).
On the other hand the variance of $m_t^2$ increases even more,
so for a system thermalized in an initial volume of 10 $fm^3$ about
$96 \%$ of the individual trajectories enter into unstable
regions ($m_t^2 < 0$) for a while during a rapid one dimensional
expansion ($\tau_0=1 fm/c$). In contrast the ensemble averaged solution
in this case remains stable.
This result supports the idea of looking for disoriented chiral
condensate (DCC) formation in
individual events.
\\[2mm]
\end{abstract}
%
%

%%%%%%%%%%%%%%%%%%%%%%%%%%% TEXT %%%%%%%%%%%%%%%%%%%%%%%%%%%%%%%%%%%

%\vspace{1.0cm}
%\newpage
%\pagenumbering{arabic}
%
Ultrarelativistic heavy ion collisions offer the possibility to study
hadronic matter at high initial temperatures and energy densities.
%(up to several $GeV/fm^3$). 
Because of their small masses
pions contribute most dominantly to the large multiplicity of produced
particles. It was speculated that some of the observed pions are
produced by the coherent decay of a semiclassical pion field
\cite{An89,Bl92}. The idea of so called `{\em disoriented chiral condensates}'
(DCC) was then made widely known due to Bjorken, Kowalski and Taylor
\cite{Bj93}. They speculated that events can occur in which the classical
pion field is oriented along a single direction in isospin space.

Rajagopal and Wilczek \cite{Ra93}
have proposed that in a rapid quench through the second order chiral QCD
phase transition such disordered chiral configurations may emerge.
%triggering a considerable research activity in recent years.
Unstable, exponentially increasing long wavelength pion fields can develop
during the `roll down' period of the order parameter according to the
pure classical equation of motion
\cite{Ra93,As95}. The spontaneous growth and subsequent
decay of these configurations would give rise to large collective fluctuations
in the number of produced neutral pions compared to charged pions,
and thus could provide a mechanism explaining a family of peculiar
cosmic ray events, the Centauros \cite{La80}. A deeper reason for these
strong fluctuations lies in the fact that all pions are assumed to sit in the
same momentum state and the overall wavefunction can carry no isospin
\cite{Gr93}. 

However, the proposed quench scenario assumes a highly
non-equilibrium initial configuration for the phase transition to occur,
namely that the potential governing the evolution of the order
parameter and the long wavelength modes immediately turns to the classical
one governing the vacuum structure at zero temperature. This represents
a very drastic assumption as the soft and classical modes completely
decouple in an ad hoc manner from the residual thermal fluctuations 
%(or hard particles)
at or slightly below the critical temperature. It is not likely
to happen in an ultrarelativistic heavy ion collision.

An alternative scenario was suggested by Gavin and M\"uller \cite{Ga94}
who proposed to use instead of the bare potential the
effective one-loop potential including the thermal fluctuations 
on the mean-field level. If the system cools rapidly enough
(due to longitudinal (D=1) or radial (D=3) expansion) and, similar to the
quench scenario, if the order parameter is small enough at the onset of the
evolution below the critical temperature, DCC might emerge as well.
The latter assumption, however, has also been criticized: if the
order parameter stays in thermal contact with the fluctuations
giving rise to the effective 1-loop potential, then one also has to allow for 
the thermal fluctuations in the initial conditions \cite{Ra96,Bi96}.
Quenched initial conditions seem to be statistically unlikely. 

In this letter we address the question of the `likeliness' of an
instability leading potentially to a DCC event in the light of
different initial conditions as raised by Blaizot and Krzywicki\cite{Bl96}.

In a simplified model we study the influence of the thermal modes
(`fluctuations') on the off-equilibrium evolution 
of the order parameters. These chiral fields are treated
semiclassically within the chiral 0(4) $\sigma $-model. 
The long wavelength modes, especially the zero modes 
used as order parameters, represent 
an open system which constantly interacts with the
thermal fluctuations (the `reservoir'). On the one-loop level the interaction
with the hard modes will generate an effective mass term $\frac{1}{2}\lambda
T^2$ \cite{Ga94,Ra96,Bi96}, leading to the effective Hartree-Fock potential.
It was recently shown in detail that in the $\phi ^4$-theory
hard modes can be integrated out on the
two-loop level leading to {\em dissipation} and {\em noise}
in the quasi-classical limit for the propagation of the long wavelength fields
\cite{Gr97}. The resulting equations of motion are of Langevin-type.
In the weak coupling limit, the friction coefficient $\eta $ is directly related to
the on-shell plasmon damping rate, $\eta \equiv 2 \gamma _{pl}$.
The noise term shows up as an imaginary part in the effective action and is related
to the friction via the fluctuation-dissipation theorem.
The interplay of noise and dissipation guarantees that the soft modes
eventually become thermally populated on the average.

We propose to study the following
Langevin equations of motion for the order parameters $\Phi _a =\frac{1}{V} \int d^3x \,
\phi_a ({\bf x},t)$ in a volume $V$ 
\bea
\label{EOM}
 \ddot{\Phi}_0 + \left( \frac{D}{\tau} + \eta \right) \dot{\Phi}_0 \, + \,
 m_0^2 \,  \Phi_0 &=& f_{\pi}m_{\pi}^2 + \xi_0,
\nonumber \\[2mm]
 \ddot{\Phi}_i + \left( \frac{D}{\tau} + \eta \right)\dot{\Phi}_i \, + \,
  m_0^2 \,   \Phi_i &=&  \xi_i   \, \, \, ,
\eea
with $\Phi^a = (\sigma, \pi ^1, \pi^2 , \pi ^3)$ being the
chiral meson fields and 
\be
m_0^2 = \lambda ( \FF + \frac{1}{2} T^2 - f_{\pi}^2 ) + m_{\pi}^2 .
\ee
Here $\tau$ is the proper time of the expanding system and the `dot'
denotes the derivative with respect to $\tau$. The above equations assume
a $D$-dimensional scaling expansion\cite{Ga94,Ra96,Bi96}.
%In eqs. (\ref{EOM}) we necglect modifications of the real
%part of the self-energy beyond the familiar one-loop mass modifications.

Before presenting our results some comments are in order: We use the standard parameters
$f_{\pi } = 93 $ MeV for the pion decay constant, $m_{\pi } = 140 $ MeV for
the pion mass, and $\lambda = 20 $ for the coupling constant. With this choice
we are obviously
in the strong coupling regime, so our conclusions drawn from the
investigation of the weak coupling regime \cite{Gr97} are rather
a motivation for our present usage of 
simple effective terms for  noise and dissipation.

We treat the dissipation term as Markovian,
which assumes a clear separation among
the timescales of the hard and soft modes.
In the semiclassical Markovian approximation the noise is effectively 
white and at two-loop level it is Gaussian \cite{Gr97},
\bea
\label{noise}
\langle \xi _a  (t) \rangle &=& 0 \, \, \, ,  \nonumber
\\[2mm]
\langle \xi _a  (t_1) \xi _b (t_2) \rangle &=&
\frac{2 T}{V} \eta \delta _{ab} \delta (t_1-t_2)
\, \, \, .
\eea
where $T$ is the temperature, $V$ the volume and $\eta $ 
the friction coefficient.

Finally we have to specify the friction coefficient $\eta $
for the $\sigma $ and pion field. The on-shell
($\omega = m $) plasmon damping rate for standard $\phi ^4$-theory
arising from the `sunset' diagram \cite{Par92,Gr97} can be easily taken over to the
0(4)-model, assuming that all four masses $m$ for the fluctuations
(i.e. hard quanta) are equal:
\be
\label{fric1}
\eta \, = \, 2 \gamma _{pl} \, = \,
\frac{9}{16 \pi ^3} \lambda ^2  \frac{T^2}{m} \, f_{Sp} (1-e^{-\frac{m}{T}})
\, \, \, ,
\ee
where $f_{Sp}(x) = -\int_{1}^{x} dt \frac{\ln t}{t-1} $ defines the
Spence function. Admittingly this `choice' is only a crude estimate as
the zero modes do not evolve on-shell during the (possibly unstable)
evolution. Thus the dissipation and noise correlation should better
be described by nonmarkovian terms including memory effects. In addition,
the O(4) transverse and longitudinal mass for the fluctuations
\cite{Ga94,Ra96,Bi96}, 
\bea
\label{masses}
m_t^2 &=& \lambda ( \FF + \frac{1}{2} T^2 - f_{\pi}^2 ) + m_{\pi}^2 \, \, \, ,
\nonumber \\[2mm]
m_l^2 &=& m_t^2 + 2\lambda (\FF ) \, \, \, ,
\eea
are not really equal. 

The phenomenon of long wavelength DCC amplification
occurs in periods when the transverse mass squared $m_t^2$ becomes negative.
This happens during a quench scenario or rapid cooling. 
For the sake of simplicity we treat $\eta $ as a constant, obtained using
(\ref{fric1}) and $m/T\approx 1$ throughout the evolution.
%\be
%\label{fric2}
%\eta \, =  \,
%\frac{9}{16 \pi ^3} \lambda ^2  T \, f_{Sp} (1-e^{-1})
%\, \, \, .
%\ee
At $T=T_c\equiv \sqrt{2 f_{\pi}^2 - 2m_{\pi }^2/\lambda } = 123$ MeV
the friction $\eta=2\gamma_{pl}=2.2$ (fm/c)$^{-1}$
is rather strong. Therefore we also investigate scenarios with
$1/2$ and $1/4$ of this value.

Aside from a theoretical justification one can  regard the Langevin
equation as a practical tool to study the effect of thermalization
on a subsystem, to sample a large set of possible trajectories
in the evolution, and to address also the question of all thermodynamically
possible initial configurations in a systematic manner.
Applying this we are able to study the up to now unknown influence of thermal
fluctuations on the growth of disoriented chiral domains.
The friction coefficient $\eta $ helps to prolong the evolution
of the order parameter, as it slows down the rolling down of the
effective potential. On the other hand the noise term leads to a
subsequent thermalization making DCC formation less likely, but allowing
also for fluctuations
on an event by event basis.  Thus we need to study
how fast the system cools and destabilizes and how fast it can thermalize.

The scaling expansion and cooling of the system is described by the equations
\be
  \frac{\dot{T}}{T} + \frac{D}{3\tau} = 0, 
\qquad   \frac{\dot{V}}{V} - \frac{D}{\tau} = 0. 
\ee

In the framework of the model described so far we investigate
different evolution scenarios for the order parameters $\Phi_a$.
For comparison we calculate the pure classical ($\eta=0, T=0$)
and the one-loop annealing scenario ($\eta=0$) 
for the $k=0$ modes only with quenched initial
condition \hbox{($\Phi_a = 0, \dot{\Phi}_a = 0, a=0 \ldots 3$).} 
These scenarios contain long and highly unstable periods
of the evolution during a rapid ($\tau_0=1 fm/c$) one-dimesional
($D=1$) scaling expansion. This can be inspected in the lower part
of Fig.1 where the quantity \hbox{$\mu_t$=sign$(m_t^2)\sqrt{|m_t^2|}$} 
is plotted as a function of time. 

In the two-loop motivated Langevin scenario we let the system
thermalize at temperature $T=T_c$ for $10$ fm/c from the quenched
initial condition
and then switch on the one-dimesional
expansion. The outcome of thermalization depends on the volume
occupied by the $k=0$ modes -- in agreement with the equipartition theorem.
The middle part of Fig.1 shows the average evolution of $1000$
trajectories, each propagated according to the Langevin equation (\ref{EOM}), 
in an initial volume of $V_0=1, 10, 100$ and $1000$ fm$^3$.
Because of the high friction coefficient $\eta$ 
the equipartition is set after the first few fm/c. 
We find that the smaller the volume, the larger $m_t^2$.
The volume dependence of the thermalized chiral
order parameter fields can be analyzed using the virial theorem
\be
\Phi_a \frac{\partial H}{\partial \Phi_a} =
\dot{\Phi}_a \frac{\partial H}{\partial \dot{\Phi}_a} = T.
\ee
In the $m_{\pi}=0$ limit analytical results can be obtained as
\be
\sum_{a=0}^3 \Phi^2_a = \frac{f}{2} \left( 1 \pm \sqrt{1+16T/\lambda Vf^2} \right)
\ee
with $f=\frac{1}{2}(T_c^2-T^2)$. The large volume limit of this
result is
\be
\sum_{a=0}^3 \Phi^2_a = \left\{ \begin{array}{lr} f + \frac{4T}{\lambda Vf} + {\cal O}(1/V^2) & T < T_c \\[0.5em]  
      2\sqrt{\frac{T_c}{\lambda V} } & T=T_c
      \\[0.5em]
    \frac{4T}{\lambda Vf} + {\cal O}(1/V^2) & T>T_c  \end{array} \right. 
\ee
From this 
\hbox{$m_t^2=\lambda (\sum \Phi_a^2 - f) = {\cal O}(1/V)$}
follows for $T < T_c$
in accordance with the Goldstone theorem.
It is also enlightening to calculate the energy per chiral degree of
freedom extra to the mean field energy
\be
\frac{\langle E \rangle - E_{MF} }{T}  = \left\{ \begin{array}{lr} \frac{1}{2} + {\cal O}(1/V)& T<T_c \\[0.5em]
  \frac{3}{4} & T=T_c \\[0.5em]
  1 + {\cal O}(1/V)& T>T_c \end{array} \right. \qquad ,  
\ee
which reflects the change of the effective number of collective chiral
degrees of freedom at the phase transition.

%%%%%%%%%%%%%%%%%%%%%%%%%% FIG.1 %%%%%%%%%%%%%%%%%%%%%%%%%%%%%

\begin{figure}
\centerline{\psfig{figure=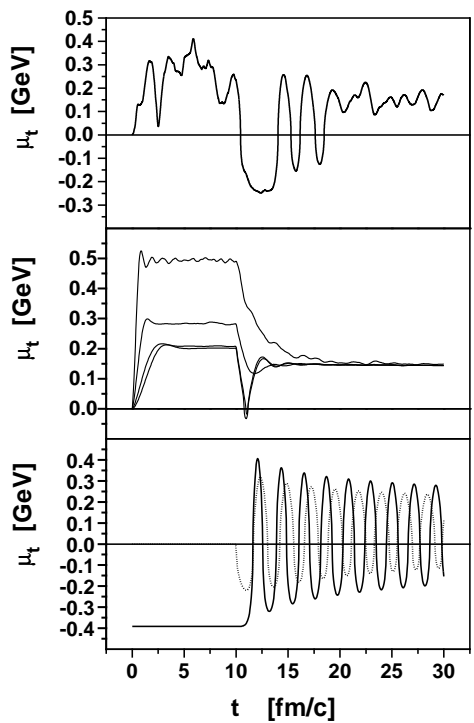,height=10cm}}
\label{Fig1}
\vspace{0.2cm}
\caption{  $\mu_t(t)$ 
    for the most unstable event of $1000$ with $V_0=10 fm^3$,
    for the average with $V_0=1, 10, 100$ and $1000 fm^3$ (from top
   to bottom), and for the pure and annealing scenarios. }
\end{figure}

%%%%%%%%%%%%%%%%%%%%%%%%%%%%%%%%%%%%%%%%%%%%%%%%%%%%%%%%%%%%%%

At the end of the expansion $\mu_t$ relaxes to the value of 
\hbox{$m_{\pi}=140$ MeV},
as can be seen from the curves in the middle part of Fig.1.
These ensemble averaged curves do not show any significant 
period of instability.
This result is in agreement with those of Randrup\cite{Ra96a} for a
one-dimensional expansion from a thermalized initial condition. 

This lack of instability in the averaged evolution does not mean,
however, that DCC formation cannot be expected in heavy ion collisions: 
Using the Langevin equation we are able to explicitely single out 
particular evolutions which are the most unstable. 
The upper part of Fig.1 presents such an evolution
preheated in a $V_0=10$ fm$^3$ initial volume at $T=T_c$.
Here, inspite the initial thermalization and ongoing noise during
the expansion, quite significant unstable periods develop.
For quantifying the strength of instability we define the quantity
\cite{Ra96a}
\be
G = \int \, |m_t| \Theta(-m_t^2) \, dt.
\ee
The amplification of small amplitude instabilites with $k=0$ is then
$\exp G$.
For the particular event shown in Fig.1 $G=4.7371$.
The distribution of this quantity, $P(G)$, is shown in Fig.2 for
an initial volume of $V_0 = 10$ fm$^3$, the $1000$ individual trajectories
are all preheated to $T=T_c$.

%%%%%%%%%%%%%%%%%%%% FIG.2 %%%%%%%%%%%%%%%%%%%%%%%%%%%

\begin{figure}
\centerline{\psfig{figure=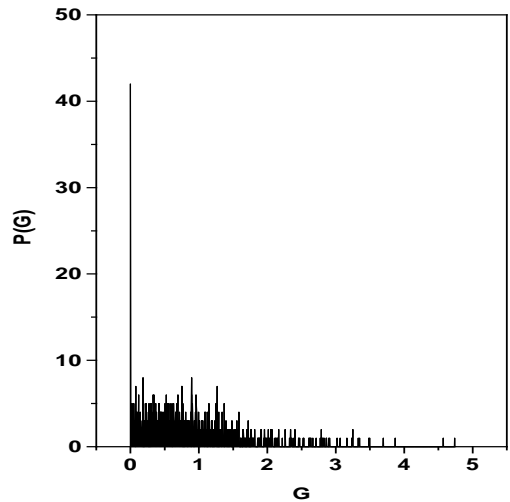,height=8cm,width=8cm}}
\label{Fig2}
\vspace{0.2cm}
\caption{ \hbox{ Distribution of the enhancement factor}
	$G$ for $V_0 = 10 fm^3$.
  	The most unstable event has $G=4.7371$.}
\end{figure}

%%%%%%%%%%%%%%%%%%%%%%%%%%%%%%%%%%%%%%%%%%%%%%%%%%%%%%%%%

We observe that only in $4.3 \%$ of all cases an unstable period is missing.
The average trajectory shows, however, no signal, because the
unstable periods occur at different times.
Therefore the middle part of Fig.1 does {\em not} allow for drawing a
conclusion. 

In order to review the tendencies Table 1 summerizes the most
important properties of the distribution of the amplification factor
in the Langevin scenario. 
The fraction of events with no instability at all are
written in the second column for the various initial volumes. 
In these cases no DCC signal can develop at all. 
The average growth factor $\langle G \rangle$
and its maximum within $1000$ events, $G_{{\rm max}}$, are shown in the
third and fourth columns, respectively.

In the late phase of the expansion the friction (\ref{fric1})
is overestimated by employing $m/T=1$,
as $m/T$ significantly increases due to the cooling
and the thermal fluctuations decouple.
In fact applying a smaller friction
the unstable oscillating periods appear for a longer time
similar to the annealing scenario.
On the other hand a variation of the friction parameter, 
$\eta$, to its $1/2$ or $1/4$
value does not change significantly the characteristics of
the distribution $P(G)$ (cf. Table.1). 

%\vspace{0.5cm}
In conclusion we have investigated the chiral O(4) model in a 
Langevin scenario,
motivated by the two-loop results of the non-equilibrium field theory.
In this approach the $k=0$ order parameter fields are
coupled to a thermal bath via mass, friction and noise terms.
We preheated the system at $T=T_c$ 
using its own dynamics and then switched over to
a one-dimensional scaling expansion.
Average and statistical properties of individual solutions of the
Langevin equations were studied with the emphasis on such periods of
the time evolution when the transverse mass becomes imaginary
and therefore an exponential growth of unstable fluctations in the
collective fields can be expected.

We have found that in different realistic initial volumes
ranging from $1$ to $1000$ fm$^3$, where the average evolution does not show
any sensible instability, individual events lead to sometimes significant growth
of fluctuations. Both the strength ($G \approx 2 - 5$) and the probability
($50 - 90 \%$) of such events are remarkably high.
The most extreme events are quite similar
to the predictions of the quenched models.
All these estimates are optimistic, because the back reaction
of the soft but $k\ne 0$ modes will lead to a shortening of 
the unstable periods \cite{Boy95}.

Our findings support the idea of looking for DCC formation experimentally
in individual events.

%%%%%%%%%%%%%%%%%%%%%% TABLE 1 %%%%%%%%%%%%%%%%%%%%%%%%%%%%%%%%%%%

\begin{table*}[t] 
\begin{center}
\begin{tabular}{||r||c|c|c||}
%\hline\hline
 $V$ & $P(G=0)$ & $\langle G \rangle$ & $ G_{max}$ \\ \hline\hline
 1 & 0 \% & 3.0983 & 7.1163  \\ \hline
 10 & 4.2 \% & 0.8540 & 4.7371  \\ \hline
 100 & 40.3 \% & 0.2728 & 2.0309  \\ \hline
 1000 & 40.4 \% & 0.1082 & 0.6780  \\ \hline\hline

 1 & 0 \% & 3.0883 & 8.5129  \\ \hline
 10 & 7.0 \% & 0.7684 & 4.5744  \\ \hline
 100 & 48.4 \% & 0.2011 & 1.8699  \\ \hline
 1000 & 56.5 \% & 0.0557 & 0.5327  \\ \hline\hline

 1 & 0 \% & 3.1088 & 9.0965  \\ \hline
 10 & 12.3 \% & 0.7775 & 5.6462  \\ \hline
 100 & 54.4 \% & 0.1620 & 1.8172  \\ \hline
 1000 & 59.7 \% & 0.0483 & 0.5063  %\\ \hline\hline

\end{tabular} 
\vspace{0.3cm}
\caption{Statistical properties of individual evolutions according
to the Langevin scenario. The upper table belongs to the friction
$\eta = 2 \gamma_{pl}$, the middle to $\eta = \gamma_{pl}$ and the lower
one to $\eta = \gamma_{pl}/2$.}
\end{center}
\end{table*}

%%%%%%%%%%%%%%%%%%% ACKNOWLEDGEMENT %%%%%%%%%%%%%%%%%%%%%%%%%%%%%%%%%%%%%
 
\vspace{0.5cm}
{\bf Acknowledgements:}
%\vspace{0.5cm}
This work has been supported by the Deutsche Forschungsgemeinschaft
(DFG) and the Hungarian Academy of Sciences (MTA) in the framework
of the project 79/1995  and by the Hungarian National Research Fund
(OTKA) under the project T019700.
T.S.B. thanks for the support from Graduiertenkolleg 
(Universit\"at Frankfurt/Giessen)
and for enlightening discussions with Prof. Walter Greiner.
C.G. thanks X.N.Wang for fruitful discussions.
 
%%%%%%%%%%%%%%%%% REFERENCES %%%%%%%%%%%%%%%%%%%%%%%

%%
%\newpage
\parskip0mm
\par
\footnotetext[1]{e-mail address: tsbiro@sunserv.kfki.hu}
\footnotetext[2]{e-mail address: greiner@theorie.physik.uni-giessen.de}
%
%
%----------------------------------------------------------------------%
%                             End of Text                              %
%----------------------------------------------------------------------%

\end{document}